\documentclass[prb,twocolumn,superscriptaddress,showpacs,amsmath,amssymb]{revtex4}
\usepackage{amsfonts}
\usepackage{bbm}
\usepackage{graphicx}
\usepackage{dcolumn}
\usepackage{bm}
\usepackage{subfigure}
\usepackage{color}

\begin{document}
\title{Field-free Josephson diode effect in altermagnet/normal metal/altermagnet junctions}
\author{Qiang Cheng}
\email[]{chengqiang07@mails.ucas.ac.cn}
\affiliation{School of Science, Qingdao University of Technology, Qingdao, Shandong 266520, China}
\affiliation{International Center for Quantum Materials, School of Physics, Peking University, Beijing 100871, China}

\author{Yue Mao}
\affiliation{International Center for Quantum Materials, School of Physics, Peking University, Beijing 100871, China}

\author{Qing-Feng Sun}
\email[]{sunqf@pku.edu.cn}
\affiliation{International Center for Quantum Materials, School of Physics, Peking University, Beijing 100871, China}
\affiliation{Hefei National Laboratory, Hefei 230088, China}

\begin{abstract}
The field-free and highly efficient diodes with the nonreciprocity of supercurrent are believed to be the core block of
the superconducting computing devices without dissipation.
In this paper, we propose a Josephson diode based upon altermagnets with the vanishing net macroscopic magnetization.
The nonreciprocity of supercurrent can be realized without applying any external magnetic field or ferromagnetic exchange field, which can avoid the magnetic cross-talk between the basic elements of the devices.
The high efficiency exceeding $40\%$ can be obtained and the efficiency shows the high stability when the structure parameters are changed.
The diode efficiency is antisymmetric about the relative orientation angle of the superconducting leads, so that its sign can easily be inverted by adjusting the relative orientation angle.
The symmetries satisfied by the current-phase difference relations and the diode efficiency are analyzed by considering the transformations of the junctions under the time-reversal, the spin-rotation and the mirror reflection operations.
The high efficiency and the high stability of the Josephson diode effect in our junctions provide the possibility for the design of the field-free dissipationless diode devices.
\end{abstract}
\maketitle

\section{\label{sec1}Introduction}

The nonreciprocity of supercurrent with the different critical currents
for opposite direction paves the way to the design of the nondissipative
digital logic, which has attracted extensive attention\cite{Ando,Lin,Narita,Ilic,Hou,Yuan,JJHe,Daido,Legg,Zinkl,Picoli,Hosur,Daidoprl,Kealhofer,Chazono,Yun,Karabassov,JJHe2,Chahid}.
In particular, the nonreciprocity of supercurrent in the Josephson junctions,
i.e., the Josephson diode effect (JDE), has stimulated growing interest due to the design and construction flexibility of the junctions.
The experimental observations of JDE have been reported recently\cite{Wu,Baumgartner,Banerjee,Ciaccia,Merida,Trahms,Pal,Matsuo}.
In the NbSe$_2$/Nb$_3$Br$_8$/NbSe$_2$ junctions, the stable half-wave rectification of a square-wave excitation has been realized using a very low switching current\cite{Wu}. Other involved structures include the Josephson junctions
based on the InAs quantum wells\cite{Baumgartner,Banerjee,Ciaccia}, the type-$\text{\uppercase\expandafter{\romannumeral2}}$ Dirac semimetal\cite{Pal}, the magic-angle twisted bilayer graphene\cite{Merida} and a single magnetic atom\cite{Trahms}. The non-locally controlled JDE is also detected experimentally in the Josephson junctions with the short-range coherent coupling\cite{Matsuo}.

Besides these experiments, various of theoretical schemes for JDE have been proposed\cite{Hu,Kopasov,Misaki,Halterman,Davydova,Jiang,Fominov,Zhang,Souto,Tanaka,Kokkeler,Steiner,Maiani,JXHu,Lu,Costa,Pillet,Hess,Hodt,
Cheng,Gupta,YFSun}. For example, a metallic nanowire on top of two superconducting slabs can host the nonreciprocal supercurrent when the finite-momentum Cooper paring is induced by the screening currents\cite{Davydova}.
The JDE with high efficiency is predicted in the supercurrent interferometers composed of two Josephson junctions\cite{Souto}.
The tunable JDE is theoretically shown on the surface of a topological insulator with an applied magnetic field\cite{Lu}.
The minimal theoretical model for JDE is formulated in the two-dimensional electron gas with the ferromagnetic barrier\cite{Costa}.
The controllable JDE by the magnetic configuration of the barriers is also studied in the Andreev molecules\cite{Hodt}.
Although JDE has been examined in various of Josephson structures experimentally and theoretically, an external magnetic field or the exchange field from ferromagnet is always required for most of the existing studies to cause the time-reversal symmetry breaking necessary for the formation of JDE.
This will bring the inevitable competition of magnetization and superconductivity in the Josephson junctions. Furthermore, the presence of the external magnetic field or the stray magnetic field produced by ferromagnet will cause the magnetic cross-talk between the basic elements of the logic device\cite{Jungwirth}.
The magnetic cross-talk is a negative factor for the realistic applications of JDE, which needs to be avoided, especially in the design of the high-density device\cite{JMHu}.

Recently, altermagnet (AM) as the third magnetic phase has become a hot topic in condensed matter physics\cite{Mazin,Smejkal,Smejkal2,Smejkal3,Krempasky,Lee,Zhou,Bai,QCheng}.
AM breaks the time-reversal symmetry similar to ferromagnet but it possesses the vanishing macroscopic net magnetization due to the alternating order of the magnetic moments in both the direct space and the momentum space\cite{Smejkal,Smejkal2,Smejkal3}[see Fig.{\ref{fig1}}(a)].
The absence of the macroscopic net magnetization in AM prevents the emergence of the stray magnetic field which widely presents in the structures including ferromagnet.
This makes AM an ideal magnetic material in the design and construction of heterojunctions consisting of AM and superconductor (SC).
The Andreev reflection\cite{CSun,Papaj}, the dc Josephson effect\cite{QCheng,Ouassou,Beenakker} and the Majorana corner modes\cite{YXLi,Zhu} in the heterostructures comprising AM and SC have been recently studied based on the zero net magnetization of AM. Especially, the crystalline orientation dependent transport properties are clarified in Ref.[\onlinecite{CSun,Papaj,QCheng,Ouassou,Beenakker}], which originate from the anisotropic spin polarization in AM.

In this paper, we propose a theoretical scheme for JDE in the junctions with a normal metal (NM) sandwiched between two AM-based SCs (AMSCs), i.e., the AMSC/NM/AMSC Josephson junctions as shown in Fig.\ref{fig1}(d).
The JDE in our junctions is derived from the simultaneous breaking of the time-reversal symmetry and the in-plane inversion symmetry.
The former is caused by the anisotropic $d$-wave altermagnetism with the zero macroscopic net magnetization in AMSCs
while the latter is caused by the Rashba spin-orbit coupling (RSOC) and the mismatch of the crystallographic orientation in AMSCs.
Any external magnetic field or ferromagnetic exchange field is not needed for the nonreciprocity of supercurrent in our junctions and the emergence of the stray field is completely avoided.
The high diode efficiency is achieved and can be maintained in a large range of the parameters of the junctions.
The efficiency exhibits the antisymmetric property about the relative orientation angle of AMSCs and its sign can also be inverted by altering the magnitudes of altermagnetism and RSOC.
We also analyze the current-phase difference relations (CPRs) and the diode efficiency from the viewpoint of symmetry through considering the transformations of the junctions under the time-reversal, the spin-rotation and the mirror-reflection as well as their joint operations.
The results obtained in this work will be helpful in the practical applications of the field-free superconductor diode with the low energy consumption.

\section{\label{sec2}Model and Formulation}

\begin{figure}[!htb]
\centerline{\includegraphics[width=1\columnwidth]{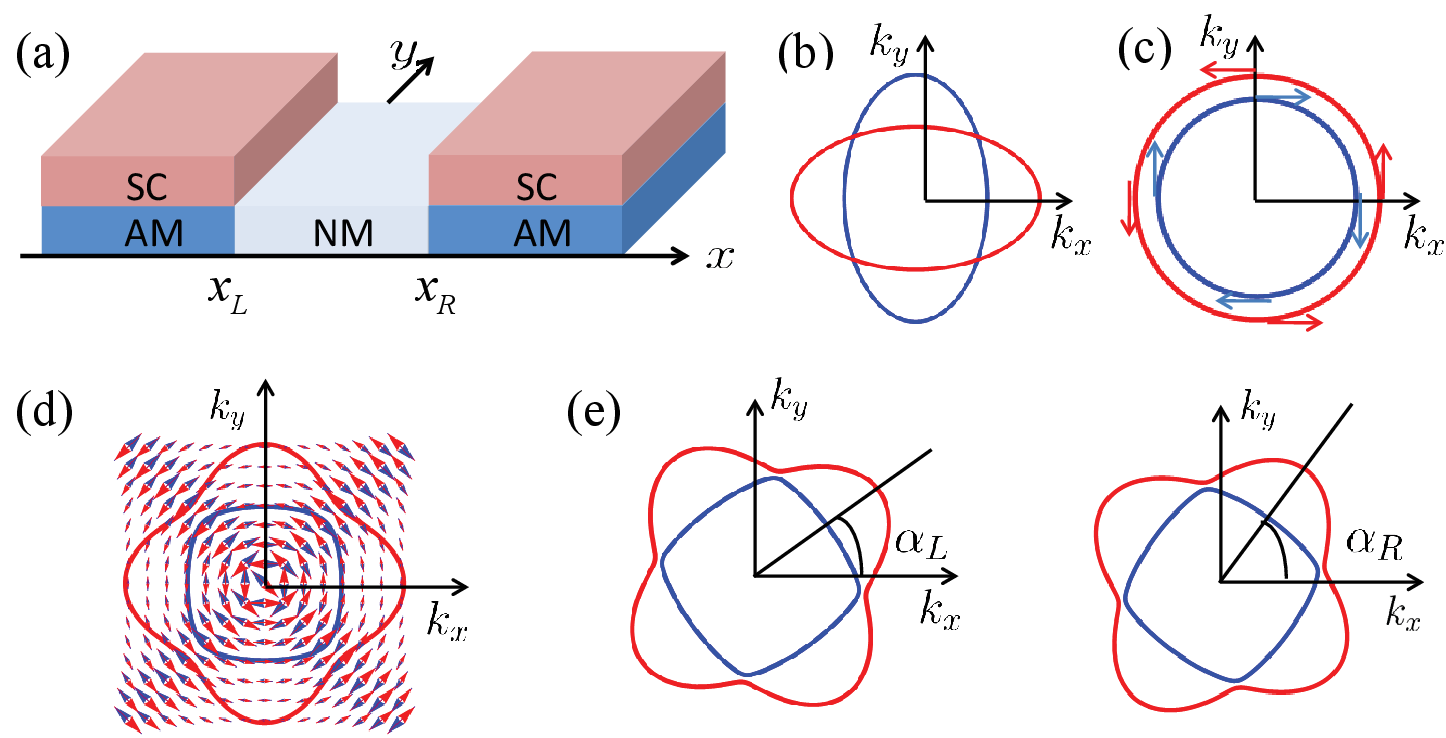}}
\caption{(a) The schematic illustration of the AMSC/NM/AMSC Josephson junctions based upon the $d$-wave AMs. (b) The elliptical Fermi surfaces of the normal state of AMSCs with altermagnetism but without RSOC. The blue (red) one is occupied by the spin-up (down) electrons. (c) The Fermi surfaces of normal state of AMSCs with RSOC but without altermagnetism. The blue and red arrows denote the spin of electrons under the spin-momentum locking. (d) The Fermi surfaces for the normal state of AMSCs with both altermagnetism and RSOC. The averages of the in-plane spin of electrons are denoted by the arrows. (e) The Fermi surfaces for the normal state of the left (right) AMSC with the orientation angle $\alpha_{l(r)}$.}\label{fig1}
\end{figure}

The AMSC/NM/AMSC Josephson junctions based upon the $d$-wave AMs in the $xy$ plane considered by us are schematically shown in Fig.\ref{fig1}(a).
The left and right semi-infinite AMSCs are formed through coupling AMs to bulk SCs with the circular Fermi surface and the isotropic $s$-wave pairing. Besides altermagnetism, RSOC in AMSCs is assumed due to the breaking of the out-plane mirror reflection. 
Before proceeding further, we now comprehensively discuss the electric structures of the normal state of AMSCs for different situations.
The Hamiltonian for the normal state of AMSCs is shown in Eq.(\ref{HAMR})in Appendix A.
For the normal state of AMSCs with altermagnetism but without RSOC,
the Fermi surfaces are two ellipses which are occupied by electrons with the opposite spin, i.e., the out-plane spin-up and the out-plane spin-down as shown in Fig.{\ref{fig1}}(b).
For the normal state of AMSCs with RSOC but without altermagnetism,
the Fermi surfaces are two concentric circles with the in-plane spin-momentum locking as shown in Fig.{\ref{fig1}}(c).
For the coexistence of altermagnetism and RSOC, the Fermi surfaces for the normal state are two closed curves as shown in Fig.{\ref{fig1}}(d).
In this situation, the spin of electrons will possess both the in-plane and the out-plane components. The blue and red arrows denoting the averages of the in-plane spin of electrons in the momentum space are calculated from the eigenvector on the Fermi surface with the same color as shown in Fig.{\ref{fig1}}(d). The calculation details can be found in Appendix A.

In this paper, we will consider the AMSCs not only with the coexistence of altermagnetism and RSOC but also with different orientation angles. The orientation of the left (right) AMSC is characterized by $\alpha_{l(r)}$ which is defined as the angle between the crystalline axis of the left (right) AMSC and the interface normal [see Fig.{\ref{fig1}}(e)].
The Fermi surfaces for the normal states of the left and
right AMSCs are schematically shown in Fig.{\ref{fig1}}(e), which can be obtained through rotating the Fermi surfaces in Fig.{\ref{fig1}}(d) about the $z$ axis by the $\alpha_{L}$ angle and the $\alpha_R$ angle, respectively.
After the rotations, corresponding changes will also occur for the spin direction of electrons.
The detailed discussions on the electric structure of the normal state of AMSCs after rotation can be found in Appendix A.
The central region of the junctions is a conventional NM without both altermagnetism and RSOC.
The two interfaces for the junctions are located at $x=x_{L}$ and $x=x_{R}$,
which are parallel to the $y$ axis as shown in Fig.{\ref{fig1}}(a). The Josephson current flows along the direction parallel to the $x$ axis. Next, we will use $\Delta x$ to denote the length of NM, i.e., $\Delta x=x_{R}-x_{L}$.

The Bogoliubov-de Gennes (BdG) Hamiltonian for the left (right) AMSC in the region $x<x_{L}$ ($x>x_{R}$) is given by $H_{L(R)}=\sum_{\bf{k}}\psi_{L(R)}^{+}({\bf{k}})\check{H}_{L(R)}({\bf{k}})\psi_{L(R)}({\bf{k}})$ with $\psi_{L(R)}({\bf{k}})=(c_{L(R){\bf{k}}\uparrow},c_{L(R){\bf{k}}\downarrow},c_{L(R){-\bf{k}}\uparrow}^{+},c_{L(R){-\bf{k}}\downarrow}^{+})^{T}$ and
\begin{eqnarray}
\check{H}_{L(R)}({\bf{k}})=\check{H}_{0}({\bf{k}})+\check{H}_{J}({\bf{k}})+\check{H}_{\lambda}({\bf{k}})+\check{H}_{\Delta}({\bf{k}}),\label{cH}
\end{eqnarray}
in the particle-hole$\otimes$spin space.
The four terms in Eq.(\ref{cH}) are expressed as
\begin{eqnarray}
\begin{split}
\check{H}_{0}({\bf{k}})=&[t_{0}(k_{x}^2+k_{y}^2)-\mu]\tau_z\otimes\sigma_0,\\
\check{H}_{J}({\bf{k}})=&t_{J}[(k_{x}^2-k_{y}^2)\cos{2\alpha_{l(r)}}+2k_{x}k_{y}\sin{2\alpha_{l(r)}}]\tau_{z}\otimes\sigma_{z},\\
\check{H}_{\lambda}({\bf{k}})=&\lambda[(k_{y}\cos{\alpha_{l(r)}}-k_{x}\sin{\alpha_{l(r)}})\tau_{0}\otimes\sigma_{x}\\
&-(k_{x}\cos{\alpha_{l(r)}}+k_{y}\sin{\alpha_{l(r)}})\tau_{z}\otimes\sigma_{y}],\\
\check{H}_{\Delta}({\bf{k}})=&-\Delta_{0}[\cos{\phi_{l(r)}}\tau_{y}+\sin{\phi_{l(r)}}\tau_{x}]\otimes\sigma_{y},\label{fourterm}
\end{split}
\end{eqnarray}
which describes the kinetic energy of particles, the $d$-wave altermagnetism, the spin-orbit coupling and the superconducting gap matrix, respectively.
Here, $t_{J}$, $\lambda$ and $\Delta_{0}$ are the magnitudes of altermagnetism, RSOC and the superconducting gap. ${\bf{k}}=(k_{x},k_{y})$ is the two dimensional wavevector in the junctions plane. $\phi_{l(r)}$ is the superconducting phase of the left (right) AMSC and the phase difference will be defined as $\phi=\phi_l-\phi_r$.
We have used $\tau_x,\tau_y,\tau_z, \tau_{0}$ and $\sigma_x,\sigma_y,\sigma_z, \sigma_0$ to denote the three components of the Pauli matrices and the identity matrices in the particle-hole space and the spin space, respectively. From Eq.(\ref{fourterm}), one can find that $\hat{H}_{J}({\bf{k}})$ is proportional to $\tau_{z}\otimes\sigma_{z}$, which implies that particles and holes will experience the opposite effects of altermagnetism. In this paper, we chose the simplest $d$-wave altermagnetism as given in Eq.(\ref{fourterm}) to demonstrate JDE based upon AMs although AMs with the higher powers of $k_x$ and $k_y$, such as the $g$-wave AMs\cite{Smejkal}, also satisfy the necessary conditions for JDE.
The BdG Hamiltonian for NM in the central region $(x_{L}<x<x_{R})$ can be written as $H_{N}=\sum_{\bf{k}}\psi_{N}^{+}({\bf{k}})\check{H}_{0}({\bf{k}})\psi_{N}({\bf{k}})$ with $\psi_{N}({\bf{k}})=(c_{N{\bf{k}}\uparrow},c_{N{\bf{k}}\downarrow},c_{N{-\bf{k}}\uparrow}^{+},c_{N{-\bf{k}}\downarrow}^{+})^{T}$ and the kinetic energy term $\check{H}_{0}({\bf{k}})=[t_{0}(k_{x}^2+k_{y}^2)-\mu]\tau_z\otimes\sigma_0$.

In order to calculate the Josephson current in the junctions, we can discretize the Hamiltonians $H_{L(R)}$ and $H_{N}$ on a two-dimensional square lattice with the lattice constant $a$ (see Appendix B).
The discrete Hamiltonian can be given by\cite{Cheng2}
\begin{equation}
\begin{aligned}
H=\sum_{{\bf{i}}}[\psi_{\bf{i}}^{+}\check{H}\psi_{\bf{i}}+\psi_{\bf{i}}^{+}\check{H}_{x}\psi_{{\bf{i}}+\delta_x}
+\psi_{\bf{i}}^{+}\check{H}_{y}\psi_{{\bf{i}}+\delta_y}\\
+\psi_{\bf{i}}^{+}\check{H}_{xy}\psi_{{\bf{i}}+\delta_x+\delta_y}+\psi_{\bf{i}}^{+}\check{H}_{x\bar{y}}\psi_{{\bf{i}}+\delta_x-\delta_y}+H.C.].\label{dH}
\end{aligned}
\end{equation}
Here, ${\bf{i}}=(i_x,i_y)$ denotes the position of sites in the lattice with $i_y$ being limited in $1\le i_y\le N_y$. The width $W$ of the junctions satisfies $W=(N_y-1)a$. The subscripts ${\bf{i}}+\delta_x$ and ${\bf{i}}+\delta_y$ represent the nearest neighbor sites of the ${\bf{i}}$th site along the $x$ direction and the $y$ direction, respectively. The subscripts ${\bf{i}}+\delta_x+\delta_y$ and ${\bf{i}}+\delta_x-\delta_y$ represent the next nearest neighbor sites of the ${\bf{i}}$th site. For the left (right) AMSC, one has $i_x\le0$ ($i_x\ge N_x+1$). The operator $\psi_{{\bf{i}}}$ is expressed as $\psi_{{\bf{i}}}=(\psi_{L(R){\bf{i}}\uparrow},\psi_{L(R){\bf{i}}\downarrow},\psi_{L(R){\bf{i}}\uparrow}^{+},\psi_{L(R){\bf{i}}\downarrow}^{+})^{T}$ in the left (right) AMSC. For NM, one has $1\le i_{x}\le N_{x}$ and its length $\Delta x$ satisfies $\Delta x=N_x a$. The operator $\psi_{{\bf{i}}}$ in NM is expressed as $\psi_{\bf{i}}=(\psi_{N\bf{i}\uparrow},\psi_{N\bf{i}\downarrow},\psi_{N\bf{i}\uparrow}^{+},\psi_{N\bf{i}\downarrow}^{+})$. The explicit expressions of the matrices $\check{H}$, $\check{H}_{x}$, $\check{H}_{y}$, $\check{H}_{xy}$ and $\check{H}_{x\bar{y}}$ for the left AMSC, the right AMSC and the central NM can be found in Appendix B.

The Hamiltonian describing the hopping between different regions can be given by
\begin{eqnarray}
H_{T} &= &\sum_{1\le i_y\le N_y}\left[\psi_{(0,i_y)}^{+}\check{T}_{L}\psi_{(1,i_y)} \right.\nonumber\\
&& \left.+\psi_{(N_x+1,i_y)}^{+}\check{T}_{R}\psi_{(N_x,i_y)}+H.C.\right],\label{hH}
\end{eqnarray}
with the hopping matrix $\check{T}_{L(R)}=\text{diag}(te^{-i\phi_{l(r)}/2},te^{-i\phi_{l(r)}/2},-te^{i\phi_{l(r)}/2},-te^{i\phi_l(r)/2})$. When we write Eqs. (\ref{dH}) and ({\ref{hH}}), a unitary transformation has been performed. Then, the superconducting phases $\phi_{l}$ and $\phi_r$ for the left and right AMSCs in Eq.(\ref{dH}) have been eliminated and they will appear in the hopping matrix $\check{T}_{L(R)}$ in Eq.(\ref{hH}). The details for the transformation can be found in Appendix B.

The particle number operator in the left AMSC can be written as
\begin{eqnarray}
N=\sum_{i_x\le 0,i_y}\sum_{\sigma=\uparrow,\downarrow}\psi_{\bf{i}\sigma}^{+}\psi_{\bf{i}\sigma}.
\end{eqnarray}
Using the Green function method, the Josephson current can be expressed as\cite{Cheng2,QFSun,addr1}
\begin{equation}
\begin{aligned}
I&=e\langle \frac{dN}{dt}\rangle\\
&=-\frac{e}{2\pi}\int dE \text{Tr}[\Gamma_z\check{T}_{L}G_{NL}^{<}(E)+H.C.],\label{JC}
\end{aligned}
\end{equation}
with $\Gamma_{z}=\sigma_z\otimes 1_{2\times2}$. From the fluctuation-dissipation theorem, the lesser Green function $G_{NL}^{<}(E)$ can be expressed as $G_{NL}^{<}(E)=-f(E)[G_{NL}^{r}(E)-G_{NL}^{a}(E)]$ with the Fermi distribution function $f(E)$. The retarded Green function $G_{NL}^{r}(E)$ and the advanced Green function $G_{NL}^{a}(E)$ can be derived from the matrices $\check{H}$, $\check{H}_x$, $\check{H}_y$, $\check{H}_{xy}$ and $\check{H}_{x\bar{y}}$ in Eq.(\ref{dH}). The details of the derivation process are presented in Appendix B.

\section{\label{sec3}Numerical results and discussions}

\begin{figure}[!htb]
\centerline{\includegraphics[width=\columnwidth]{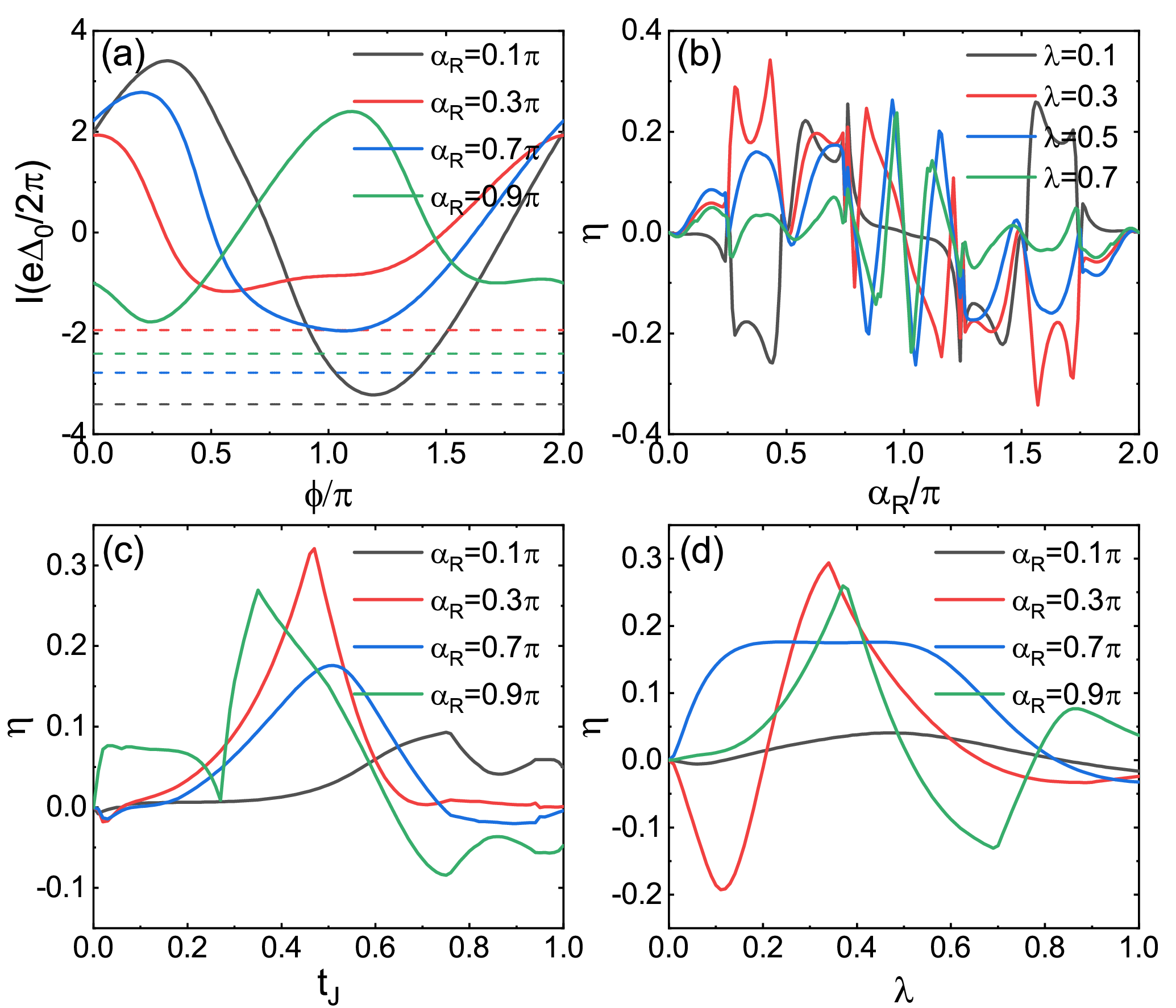}}
\caption{(a) The CPRs for different orientation angles with $t_{J}=0.5$ and $\lambda=0.3$. Each dashed line gives the critical current $-I_{C+}$ of the CPR with the same color.
(b) The variations of the diode efficiency as the orientation angle for $t_{J}=0.5$. (c) The variations of the diode efficiency as the magnitude of altermagnetism for $\lambda=0.3$. (d) The variations of the diode efficiency as RSOC strength $\lambda$ for $t_{J}=0.5$.
Other parameters are taken as $t_0=1$, $t=1$, $\mu=2$, $a=1$, $N_x=1$ and $N_y=3$.}\label{fig2}
\end{figure}

Before presenting the numerical results, we give the necessary conditions responsible for JDE in the AMSC/NM/AMSC junctions from the symmetry analysis (see Appendix C for details)\cite{Cheng3}.
For $t_{J}=0$ and $\lambda\ne 0$, there is only RSOC in AMSCs and altermagnetism is absent.
The junctions will respect the time-reversal symmetry. The continuum Hamiltonian for the left (right) AMSC can be transformed according to $\mathcal{T}H_{L(R)}(\alpha_{l(r)},\phi_{l(r)})\mathcal{T}^{-1}=H_{L(R)}(\alpha_{l(r)},-\phi_{l(r)})$ by the time-reversal operator $\mathcal{T}$. Since the time-reversal operation will invert the direction of the Josephson current, we have the following relation for CPRs,
\begin{eqnarray}
I(\alpha_l,\alpha_r,\phi)=-I(\alpha_l,\alpha_r,-\phi).\label{rel1}
\end{eqnarray}
This relation means that the positive current and the negative current possess the same critical value, and the JDE can not be expected while $t_{J}=0$.
On the other hand, for $t_{J}\ne 0$ and $\lambda=0$,
there is only altermagnetism in AMSCs and RSOC is absent.
The time-reversal symmetry will be broken due to the presence of altermagnetism. However, the symmetry under the joint operation $\mathcal{X}=\mathcal{R}_{y}(\pi)\mathcal{T}$ will be satisfied by AMSCs with $\mathcal{R}_{y}(\pi)$ being the spin-rotation about the $y$ axis by a $\pi$ angle.
The Hamiltonian for the left (right) AMSC will be transformed according to $\mathcal{X}H_{L(R)}(\alpha_{l(r)},\phi_{l(r)})\mathcal{X}^{-1}=H_{L(R)}(\alpha_{l(r)},-\phi_{l(r)})$. In this situation, the relation in Eq.(\ref{rel1}) can be obtained again and the JDE in our junctions will be forbidden. From the above discussions, the coexistence of RSOC and altermagnetism is a necessary condition for the possible emergence of the JDE in the AMSC/NM/AMSC junctions.

Another condition for the JDE is the unequal orientation angles in the left AMSC and the right AMSC.
In order to prove this point, we introduce the in-plane inversion operation $\mathcal{Y}=\mathcal{M}_{xz}\mathcal{M}_{yz}$. Here, $\mathcal{M}_{xz}$ and $\mathcal{M}_{yz}$ are the mirror reflection operations about the $xz$ plane and the $yz$ plane, respectively. The in-plane inversion operation can transform the Hamiltonian for the left (right) AMSC according to $\mathcal{Y}H_{L(R)}(\alpha_{l(r)},\phi_{l(r)})\mathcal{Y}^{-1}=H_{R(L)}(\alpha_{l(r)},\phi_{l(r)})$. The operation $\mathcal{M}_{yz}$ have exchanged the orientation angles and the superconducting phases for the left AMSC and the right AMSC. As a result, the Josephson current is inverted and the relation $I(\alpha_l,\alpha_r,\phi)=-I(\alpha_r,\alpha_l,-\phi)$ will hold. If $\alpha_l=\alpha_r=\alpha$, the relation in Eq.(\ref{rel1}) will be satisfied and the JDE will also be forbidden. In other words, the orientation angles for the left AMSC and the right AMSC must be different in order to break the in-plane inversion symmetry and produce the JDE.

Next, we will present the numerical results with the two necessary conditions for JDE being simultaneously satisfied.
In our calculations, we will fix $\alpha_l=0$ for definiteness and let $\alpha_R$ vary. Other parameters are taken as $t_{0}=1$, $\mu=2$, $t=1$,
$\Delta_0=0.01$ and $a=1$. Fig.\ref{fig2}(a) shows the CPRs for different values of $\alpha_r$ with $t_{J}=0.5$ and $\lambda=0.3$.
For each CPR curve, we define the critical value for its positive current as $I_{c+}$ with $I_{c+}=\text{max}[I(0<\phi<2\pi)]$ and the critical value for its negative current as $I_{c-}$ with $I_{c-}=\text{max}[-I(0<\phi<2\pi)]$. In Fig.\ref{fig2}(a), we plot $-I_{c+}$ of the CPRs as the dashed lines. Each dashed line and its corresponding CPR curve possess the same color.
It is found that the dashed lines are not tangent with their CPR curves. This indicates that the critical value for the positive current and that for the negative current are not equal, i.e., $I_{c+}\ne I_{c-}$.
This is just the nonreciprocity effect of the Josephson supercurrent in our junctions. Actually, all the dashed lines in Fig.\ref{fig2}(a) are below their corresponding CPRs and one has $I_{c-}<I_{c+}$. As a result, our junctions with a fixed orientation of the right AMSC can serve as an efficient rectifier by altering the direction of the applied current $I$ with $I_{c-}<I<I_{c+}$.

In order to measure the nonreciprocity in our junctions, we introduce the Josephson diode efficiency $\eta$ which is defined as
\begin{eqnarray}
\eta=\frac{I_{c+}-I_{c-}}{I_{c+}+I_{c-}}.\label{deta}
\end{eqnarray}
The diode efficiency $\eta$ as functions of $\alpha_{r}$ is shown in Fig.\ref{fig2}(b) for different values of RSOC with $t_{J}=0.5$. The efficiency $\eta$ exhibits the oscillation behavior when the orientation angle $\alpha_r$ is adjusted. However, $\eta$ can keep stable and high values in a large range of $\alpha_r$. For example, the efficiencies for $\lambda=0.1$ and $\lambda=0.3$ can reach about $20\%$ in the range of $0.25\pi<\alpha_r<0.5\pi$ and can exceed $15\%$ in the range of $0.5\pi<\alpha_r<0.75\pi$.
The stable and high efficiency with no strict demand on the value of the orientation angle in our junctions will be helpful for the practical applications of JDE based upon AMs. From Fig.\ref{fig2}(b), one can also find that raising the value of RSOC will not always increase the efficiency for the given $t_J$. For example, the efficiency in the angle range of $0.25\pi<\alpha_r<0.5\pi$ is obviously weakened when the value of RSOC is increased from $\lambda=0.3$ to $\lambda=0.5$ and then to $\lambda=0.7$. On the contrary, the smaller value $\lambda=0.1$ of RSOC can cause higher diode efficiency in the same angle range. The large value of RSOC is not the essential condition for the high diode efficiency, which is another advantage of our junctions.

In addition, the curves for the efficiency $\eta$ in Fig.\ref{fig2}(b) satisfy the following relation
\begin{eqnarray}
\eta(\alpha_r)=-\eta(2\pi-\alpha_r),\label{eta}
\end{eqnarray}
which suggests that the diode efficiency will change its sign when one adjusts the orientation angle from $\alpha_r$ to $-\alpha_r$.
In order to clarify the relation in Eq.(\ref{eta}), we introduce the joint operation $\mathcal{Z}=\mathcal{T}\mathcal{M}_{xz}$ (see Appendix C for details). The Hamiltonian $H_{L(R)}$ for the left (right) AMSC will be transformed according to $\mathcal{Z}H_{L(R)}(\alpha_{l(r)},\phi_{l(r)})\mathcal{Z}^{-1}=H_{L(R)}(2\pi-\alpha_{l(r)},-\phi_{l(r)})$ under the joint operation. As a unitary operation, the mirror reflection $\mathcal{M}_{xz}$ does not change the Josephson current. On the other hand, the time reversal operation $\mathcal{T}$ only inverts the direction of the current. Hence, we have the relation $I(\alpha_l,\alpha_r,\phi)=-I(2\pi-\alpha_l,2\pi-\alpha_r,-\phi)$ for CPRs. For $\alpha_l=0$, the relation will degenerate into $I(\alpha_r,\phi)=-I(2\pi-\alpha_r,-\phi)$. When one adjusts the orientation angle from $\alpha_r$ to $-\alpha_r$, the positive (negative) current for $\alpha_r$ will become the negative (positive) current for $-\alpha_r$. The critical value $I_{c+}(I_{c-})$ for $\alpha_r$ will turn into the critical value $I_{c-}(I_{c+})$. According to the definition of $\eta$ in Eq.(\ref{deta}), $\eta$ will change its sign when $\alpha_r$ is changed into $-\alpha_r$. Furthermore, for $\alpha_r=\pi$, we have $I(\pi,\phi)=-I(\pi,-\phi)$ and the JDE will be vanished as shown in Fig.\ref{fig2}(b).

The variations of the efficiency $\eta$ as the magnitude of altermagnetism are presented in Fig.\ref{fig2}(c) for several given values of the orientation angle $\alpha_r$. In Fig.\ref{fig2}(c), we have taken $\lambda=0.3$. For the orientation angle $\alpha_r=0.3\pi, 0.7\pi$ or $0.9\pi$, the efficiency $\eta$ reaches its peak value at the position around $t_{J}=0.4$. The peak value can exceed $15\%$, $25\%$ and even $30\%$ for $\alpha_r=0.3\pi, 0.7\pi$ and $0.9\pi$. Especially, the diode efficiency exceeding $10\%$ can be obtained in a large range of $t_{J}$ for the three orientation angles, which shows the excellent stability of the diode effect in our AMSC/NM/AMSC junctions when the strength of altermagnetism is changed. The variations of the efficiency $\eta$ as RSOC are presented in Fig.\ref{fig2}(d) for the given orientation angles and the fixed $t_J=0.5$.
The similar stability can also be found in Fig.\ref{fig2}(d) for these orientation angles when the strength of RSOC is altered.
In particular, it is found that the very small value of RSOC can bring high efficiency. For $\alpha_r=0.3\pi$ and $\alpha_r=0.7\pi$, the efficiency about $10\%$ can be achieved even for $\lambda=0.05$ which is one tenth of $t_{J}$.
In Figs.\ref{fig2}(c) and (d), we also consider the variations of $\eta$ for the small orientation angle $\alpha_r=0.1\pi$. In this situation, a larger value of altermagnetism or RSOC is needed for the appearance of the peak value of $\eta$, which corresponds to the unobvious in-plane inversion symmetry breaking for the small orientation mismatch of AMSCs. In addition, adjusting the strength of altermagnetism or RSOC can change the sign of $\eta$ as shown in Figs.\ref{fig2}(c) and (d), which can also be observed in Fig.\ref{fig2}(b).

\begin{figure}[!htb]
\centerline{\includegraphics[width=\columnwidth]{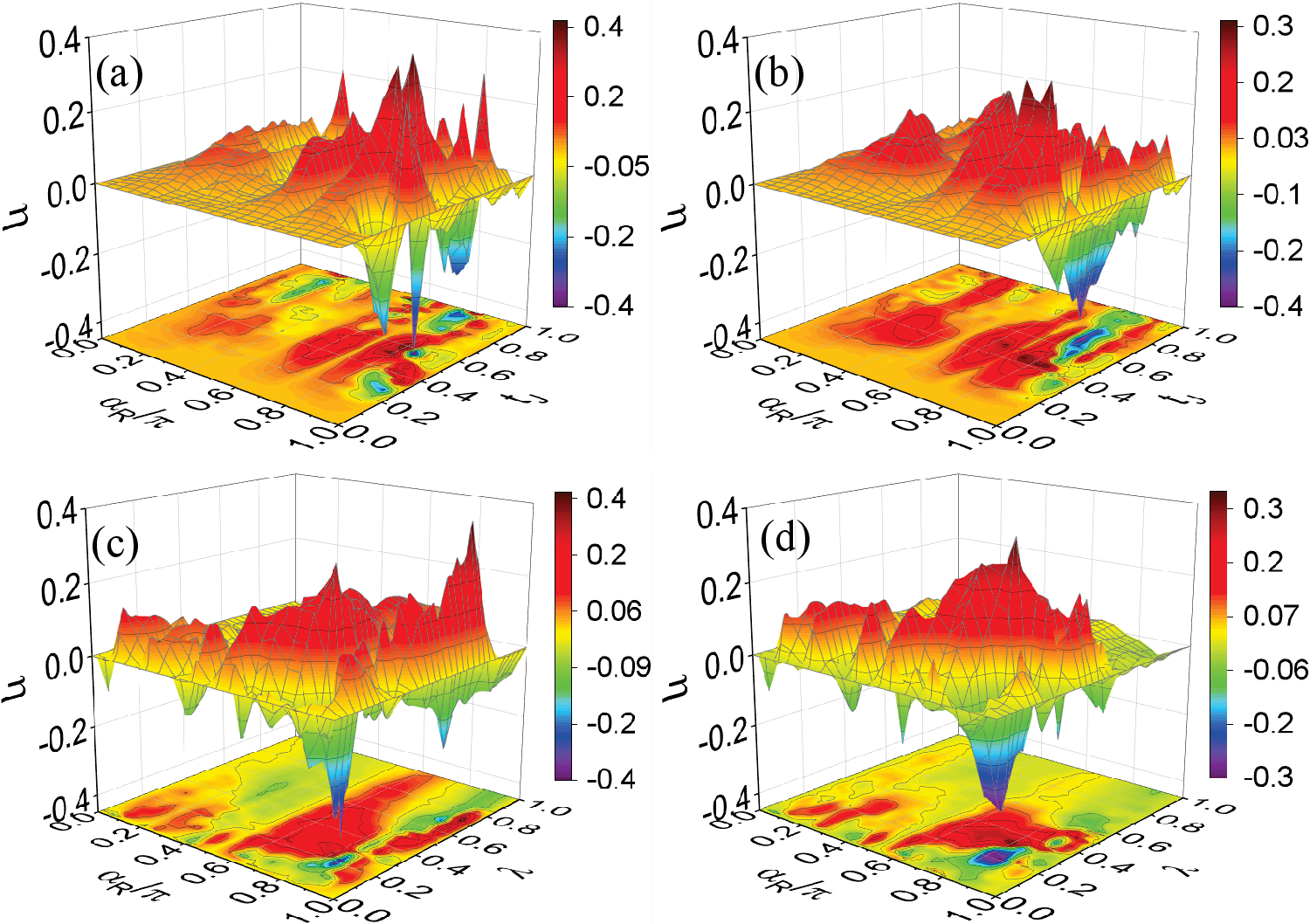}}
\caption{The Josephson diode efficiency $\eta$ for (a) $N_{x}=1$ and (b) $N_x=10$ in the $(\alpha_{R},t_{J})$ space and for (c) $N_x=1$ and $N_x=10$ in the $(\alpha_{R},\lambda)$ space. For the width of the junctions, we have taken $N_y=15$. In (a) and (b), $\lambda$ is taken as 0.3; in (c) and (d), $t_J$ is taken as 0.5.}\label{fig3}
\end{figure}

In Fig.\ref{fig2}, we have taken $N_{y}=3$ which corresponds to the narrow limit of the junctions. For the discrete model in Eq.(\ref{dH}), the next neighbor hopping is required in order to describe the altermagnetism in AMSCs. The junctions with $N_{y}=3$ are enough to seize the basic physics in our Josephson structure based upon AMs. However, for the practical applications, it is necessary to consider the influence of the width of the junctions on the JDE. In Fig.\ref{fig3}(a), we present the diode efficiency $\eta$ for $N_y=15$ as a bivariate function of $\alpha_R$ and $t_J$ and its projection on the $(\alpha_R,t_J)$ plane.
It is found that the peak value of $\eta$ can exceed $36\%$ and the valley value of $\eta$ is less than $-44\%$. The obtained high efficiency implies that an increase in width of the AMSC/NM/AMSC junctions will not weaken the JDE. From the projection of $\eta$, one can also find that the parameters for the high efficiency distribute in a large area in the $(\alpha_R,t_J)$ plane. The area is approximately centered around the point with $\alpha_R=0.8\pi$ and $t_J=0.4$ as shown in Fig.\ref{fig3}(a). The similar property of the diode efficiency can manifest itself in Fig.\ref{fig3}(c) when $\eta$ is expressed as a bivariate function of $\alpha_R$ and $\lambda$. In this situation, the peak value of $\eta$ will exceed $37\%$ and the valley value can reach $-39\%$. The parameters $\alpha_R$ and $\lambda$ for the high efficiency mainly distribute around the point with $\alpha_R=0.7\pi$ and $\lambda=0.3$ in the $(\alpha_R,\lambda)$ plane.
The above discussions on the numerical results in Fig.\ref{fig3}(a) and (c) indicate the high stability and the high efficiency of the JDE in
the AMSC/NM/AMSC junctions when the width of the junctions is increased.
Next, we discuss the influence of the length of the junctions on the JDE. In Fig.\ref{fig3}(a) and (c), we have taken $N_x=1$ which corresponds to the short limit of the junctions.
In Fig.\ref{fig3}(b) and (d), the numerical results of the efficiency $\eta$ for $N_x=10$ are presented as functions of $(\alpha_R,t_J)$ and $(\alpha_R,\lambda)$, respectively.
The increase of the length will not significantly weaken the diode efficiency. The peak value and the valley value of $\eta$ in Fig.\ref{fig3}(b) can reach about $32\%$ and $-38\%$ while those values in Fig.\ref{fig3}(d) can reach about $37\%$ and $-32\%$.
Although the distribution area of the parameters responsible for the high efficiency in \ref{fig3}(d) becomes a bit smaller, the positions of the center of the area does not change as shown in Fig.\ref{fig3}(b) and \ref{fig3}(d).
These properties of $\eta$ demonstrate that the stability and the high efficiency of the JDE can be maintained in our junctions when the length of the junctions is increased. In Fig.\ref{fig3}, the orientation angle $\alpha_R$ of the right AMSC is only taken as $0\le\alpha_R\le\pi$ since the antisymmetric relation for $\eta$ in Eq.(\ref{eta}) still holds for the wider and longer junctions with $N_y=15$ or $N_x=10$.

Here, we give some discussions on several scales and their relations in our junctions.
The first scale is the coherence length $\xi_0=\hbar v_F/\Delta_{0} = 2\sqrt{t_0\mu}/\Delta_0$ in superconductors, which can be calculated as about $283a$ using $t_0=1$, $\mu=2$ and $\Delta_0=0.01$.
The second scale is the length of NM, which is given by $\Delta x=N_{x}a$ as shown in Fig.{\ref{figA3}}. From the two scales, one can find that the short junctions are considered in this paper since we have $\Delta x\ll\xi_0$ even for $N_x=10$ in Fig.{\ref{fig3}}.
Due to the presence of AMs with RSOC in our junctions, the spin of electrons is not conserved. In this situation, both singlet and triplet superconducting correlations will arise in NM\cite{Bergeret,Kadigrobov}.
The third scale defined by us is the decay length $\xi$ of the singlet correlation in AM, which is given by $\xi=\hbar v_{F}/t_{J}$. For a moderate value of $t_{J}$ with $t_{J}=0.5$, $\xi=\xi_{0}/50\approx5.66a$. When the length of NM increases from $\Delta x=a$ with $N_x=1$ to $\Delta x=10a$ with $N_x=10$, the singlet correlation will decay fast and finally becomes negligible\cite{Bergeret,Kadigrobov}. However, the triplet correlation possesses lager penetration length and will decay slowly as the increase of $\Delta x$.\cite{Bergeret,Kadigrobov}
It will dominate the Josephson effects in our junctions with $\Delta x=10a$ and in longer junctions with $\Delta x>10a$.

Now, we discuss the experimental realization of the JDE in our AMSC/NM/AMSC junctions. Firstly, the high efficiency of the JDE can be obtained when $t_J$ is of the order of $10^{-1}t_0$, e.g. $t_J\approx0.4t_0$ in Fig.\ref{fig3} or $t_J\approx0.5t_0$ in Fig.\ref{fig2}(c), which is just the typical magnitude of altermagnetism in AM\cite{Smejkal2,Beenakker}. For RSOC, its value in the AM without coupling to SC is about $t_J/10$ (see Supplemental
Material of Ref.[\onlinecite{Smejkal2}]). The SC on AM will enhance the strength of RSOC due to the breaking of the reflection symmetry in AM/SC heterostructure. The strength of RSOC can also be enhanced by an applied electric field or the substrate beneath AM. Actually, even for $\lambda=t_J/10$, the diode efficiency exceeding $10\%$ can be acquired in our junctions as discussed above and as shown in Fig.\ref{fig2} and Fig.\ref{fig3}. Secondly, the detection of the Josephson effect in junctions composed of materials with the different crystallographic orientation is a mature technology in experiment\cite{Chesca}. It is expected that the high efficiency in our junctions can be detected for a given orientation angle of the right AMSC. Finally, we noticed that two theoretical schemes for the field-free superconducting diode have been proposed by the authors of Refs.[\onlinecite{Jiang}] and [\onlinecite{Kokkeler}]. It is worth emphasizing that the field-free diode in these works mean that the realization of the diode effect does not need an applied magnetic field. But the periodic arrangement of magnetic dots or the ferromagnetic insulator is still needed in these schemes, which implies that the magnetic cross-talk in these schemes can not be avoided. This is distinct from our field-free scheme based upon AMs in which any magnetic field is not needed and the magnetic cross-talk can be completely eliminated.

\section{\label{sec4}Conclusions}

We theoretically study the nonreciprocity of supercurrent in the Josephson junctions based upon AMs. The JDE with the high efficiency can be realized without an external field and the ferromagnetic exchange field, which keeps stable in a large range of parameter values. The sign change of the diode efficiency can be brought by rotating the orientation angle in AMSC or altering the magnitudes of altermagnetism and RSOC.
The high efficient JDE can well survive regardless of the width or the length of the junctions.
We also analyze the symmetry of the CPRs based on the transformations of the time-reversal, the spin-rotation and the mirror reflection operations, and show that the occurrence of JDE is a result of the combination of both the time-reversal symmetry breaking and the in-plane inversion breaking.
The investigations in this paper open the possibility for the study on and the application of the field-free JDE based upon AMs.

\section*{\label{sec5}ACKNOWLEDGMENTS}

This work was financially supported
by the National Natural Science Foundation of China
under Grants Nos. 12374034, 11921005 and 11447175,
the Innovation Program for Quantum Science and Technology (2021ZD0302403),
the Strategic Priority Research Program of Chinese Academy of Sciences (XDB28000000)
and the project ZR2023MA005 supported by Shandong Provincial Natural Science Foundation. We acknowledge
the Highperformance Computing Platform of Peking University
for providing computational resources.

\section*{\label{sec5} APPENDIX}

\setcounter{equation}{0}
\setcounter{figure}{0}
\renewcommand{\theequation}{A\arabic{equation}}
\renewcommand{\thefigure}{A\arabic{figure}}

\subsection{The electronic structure for the normal state of AMSCs}\label{A}

The continuum Hamiltonian for the normal state of the left or the right AMSC, i.e., the AM system with RSOC, can be written as
\begin{eqnarray}
h_{AM}=\sum_{\bf{k}}\tilde{\psi}_{L(R)}^{+}({\bf{k}})\hat{h}_{AM}({\bf{k}})
 \tilde{\psi}_{L(R)}({\bf{k}}),\label{HAMR}
\end{eqnarray}
with $\hat{h}_{AM}({\bf{k}})=\hat{h}_{0}({\bf{k}})+\hat{h}_{J}({\bf{k}})+\hat{h}_{\lambda}({\bf{k}})$ and $\tilde{\psi}_{L(R)}({\bf{k}})
=(c_{L(R){\bf{k}}\uparrow},c_{L(R){\bf{k}}\downarrow})^{T}$ in the spin space.
Here, $\hat{h}_{0}({\bf{k}})=[t_{0}(k_{x}^2+k_{y}^2)-\mu]\sigma_0$,  $\hat{h}_{J}({\bf{k}})=t_{J}[(k_{x}^2-k_{y}^2)\cos{2\alpha}+2k_{x}k_{y}\sin{2\alpha}]\sigma_{z}$ and $\hat{h}_{\lambda}({\bf{k}})=\lambda[(k_{y}\cos{\alpha}-k_{x}\sin{\alpha})\sigma_{x}
-(k_{x}\cos{\alpha}+k_{y}\sin{\alpha})\sigma_{y}]$ in the spin space of particles. The orientation angle is denoted by $\alpha$,
in which the subscript $l(r)$ is omitted for brevity.
The eigenvalue equation for the particles in the normal state can be written as
\begin{eqnarray}
\hat{h}_{AM}({\bf{k}})\chi({\bf{k}})=E\chi({\bf{k}}).
\end{eqnarray}
The energies of the particles can be solved from the eigenvalue equation as
\begin{eqnarray}
E_{\pm}=t_0(k^2-\mu)\pm\xi({\bf{k}}),\label{Epm}
\end{eqnarray}
with $\xi({\bf{k}})=\frac{1}{\sqrt{2}}\{k^2(t_J^2k^2+2\lambda^2)+t_{J}^2[(k_x^4+k_y^4-6k_x^2 k_y^2)\cos{4\alpha}+4k_x k_y(k_x^2-k_y^2)\sin{4\alpha}]\}^{1/2}$
and $k^2=k_x^2+k_y^2$.
The eigenvectors can be written correspondingly as
\begin{eqnarray}
\chi_{+(-)}({\bf{k}})=\left(\begin{array}{c}
+(-)ie^{i\alpha}\frac{[\xi({\bf{k}})+\hat{h}_{J11(22)}({\bf{k}})]}{\lambda(k_x+ik_y)}\\
1
\end{array}\right),\label{evs}
\end{eqnarray}
with $\hat{h}_{J11(22)}({\bf{k}})$ being the $11(22)$ component of the matrix $\hat{h}_{J}({\bf{k}})$.
The averages of the spin of the particles in the normal state can be calculated using the eigenvectors in Eq.(\ref{evs}), which components are given by
\begin{equation}
\begin{aligned}
&\langle s_{x+(-)}\rangle=\\
&+(-)\frac{[\xi({\bf{k}})+\hat{h}_{J11(22)}({\bf{k}})]\lambda(k_y\cos{\alpha}-k_x\sin{\alpha})}{[\xi({\bf{k}})+\hat{h}_{J11(22)}({\bf{k}})]^2+\lambda^2k^2},\\
&\langle s_{y+(-)}\rangle=\\
&-(+)\frac{[\xi({\bf{k}})+\hat{h}_{J11(22)}({\bf{k}})]\lambda(k_y\sin{\alpha}+k_x\cos{\alpha})}{[\xi({\bf{k}})+\hat{h}_{J11(22)}({\bf{k}})]^2+\lambda^2k^2},\\
&\langle s_{z+(-)}\rangle=\frac{1}{2}\frac{[\xi({\bf{k}})+\hat{h}_{J11(22)}({\bf{k}})]^2-\lambda^2k^2}{[\xi({\bf{k}})+\hat{h}_{J11(22)}({\bf{k}})]^2+\lambda^2k^2}.
\end{aligned}\label{asp}
\end{equation}
Here, $s_{x+(-)}$, $s_{y+(-)}$ and $ s_{z+(-)}$ are the $x$, $y$ and $z$ components of spin for the particles with energies $E_{+(-)}$.

\begin{figure}[!htb]
\centerline{\includegraphics[width=\columnwidth]{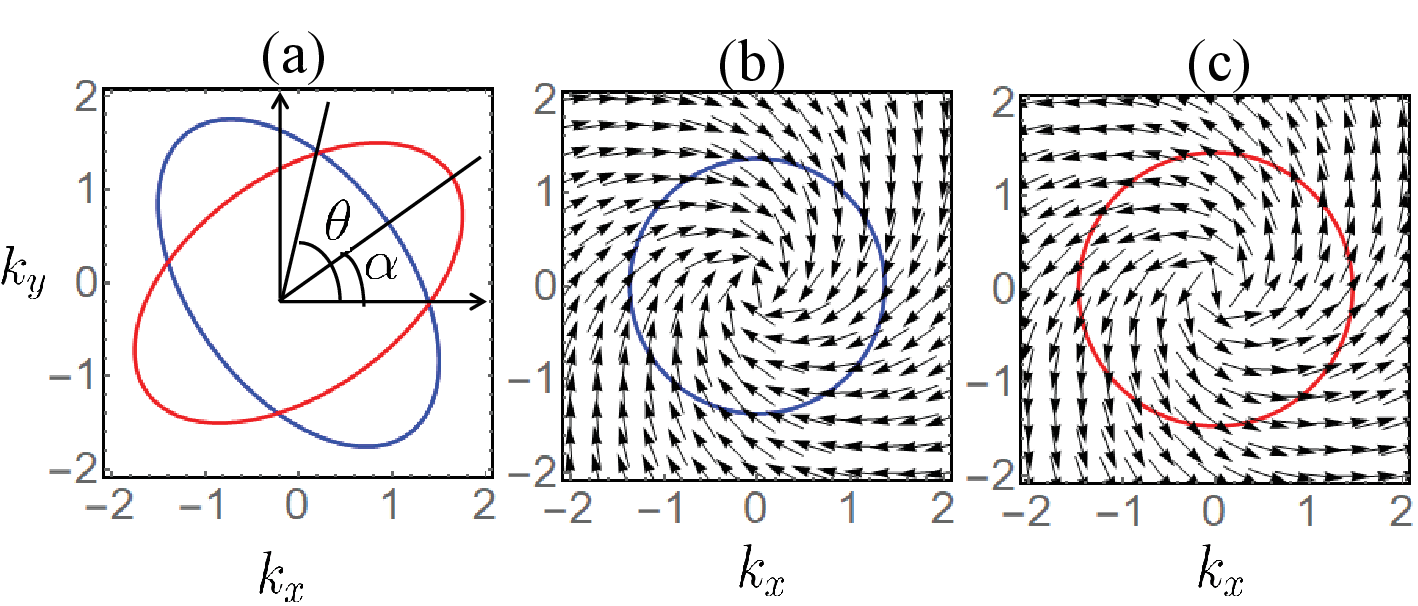}}
\caption{(a) The elliptical Fermi surfaces of AM without RSOC.
(b) and (c) are the circular Fermi surfaces for the $E_{+}$ and $E_{-}$ bands of the two-dimensional electron gas with only RSOC. The black arrows denote the spin of particles in the momentum space. The parameters are taken as $t_{0}=1,t_{J}=0.5,\lambda=0.1,\alpha=0.2\pi$ and $\mu=2$.}\label{figA1}
\end{figure}

In order to understand the electric structure of the normal state,
i.e., the AM with RSOC, we first discuss the electric structures
for the system with $\lambda \ne 0$ or $t_{J}\ne 0$ only.
For $\lambda=0$ and $t_{J}\ne 0$, the Hamiltonian in Eq.(\ref{HAMR}) describes the AM without RSOC.
In this case, the Fermi surfaces are two ellipses populated by electrons with the opposite spin due to the altermagnetism in AM\cite{Smejkal}, as shown in Fig.\ref{figA1}(a).
The electrons on the blue ellipse have the spin along the $+z$ direction while those on the red ellipse have the spin along the $-z$ direction.
The two Fermi surfaces intersect at $\theta=\alpha+(2n+1)\pi/4$ in the momentum space with $n$ being an integer number.
In particular, for the $d_{x^2-y^2}$-wave altermagnetism with $\alpha=0$,
the two Fermi surfaces intersect at $\theta=(2n+1)\pi/4$ while for the $d_{xy}$-wave altermagnetism with $\alpha=\pi/4$, the Fermi surfaces intersect at $\theta=n\pi/2$.
At the intersection points, the electrons are spin-unpolarized.
Otherwise, the electrons are spin-polarized on the Fermi surfaces.

On the other hand, for $t_{J}=0$ and $\lambda\ne 0$,
the Hamiltonian in Eq.(\ref{HAMR}) describes the usual two-dimensional electron gas with only RSOC after the $\alpha$ angle rotation of the crystallographic axis.
In this case, the Fermi surfaces are two concentric circles with different radius as shown in Figs.\ref{figA1}(b) and (c).
The blue circle is the Fermi surface determined by $E_{+}=0$
while the red circle is the Fermi surface determined by $E_{-}=0$.
The spin of particles on the two Fermi surfaces is pinned in the $xy$ plane with $\langle s_{z+(-)}\rangle=0$ due to the presence of RSOC.
The spin textures are formed on the two Fermi surfaces due to the spin-momentum locking.
The black arrows in Figs.\ref{figA1}(b) and (c) denote the spin vectors $(\langle s_{x+}\rangle,\langle s_{y+}\rangle)$ and $(\langle s_{x-}\rangle,\langle s_{y-}\rangle)$ in the whole momentum space.
From the plot of the spin vectors in Figs.\ref{figA1}(b) and (c), one can obtain the information of the spin distribution on the two Fermi surfaces. The dot product between the momentum and the corresponding spin vector on the Fermi surfaces can be given by $(\langle s_{x+(-)}\rangle,\langle s_{y+(-)}\rangle)\cdot(k_x,k_y)=-(+)\frac{k}{2}\sin{\alpha}$, which means that the angle between the momentum and the spin is $\alpha+\pi/2$ on the Fermi surface determined by $E_{+}=0$ and is $\alpha-\pi/2$ on the Fermi surface determined by $E_{-}=0$. If one takes $\alpha=0$, the momentum and the spin will be perpendicular, which corresponds to the familiar two-dimensional electron gas system without the $\alpha$ angle rotation of the crystallographic axis\cite{Costa}.

\begin{figure}[!htb]
\centerline{\includegraphics[width=0.8\columnwidth]{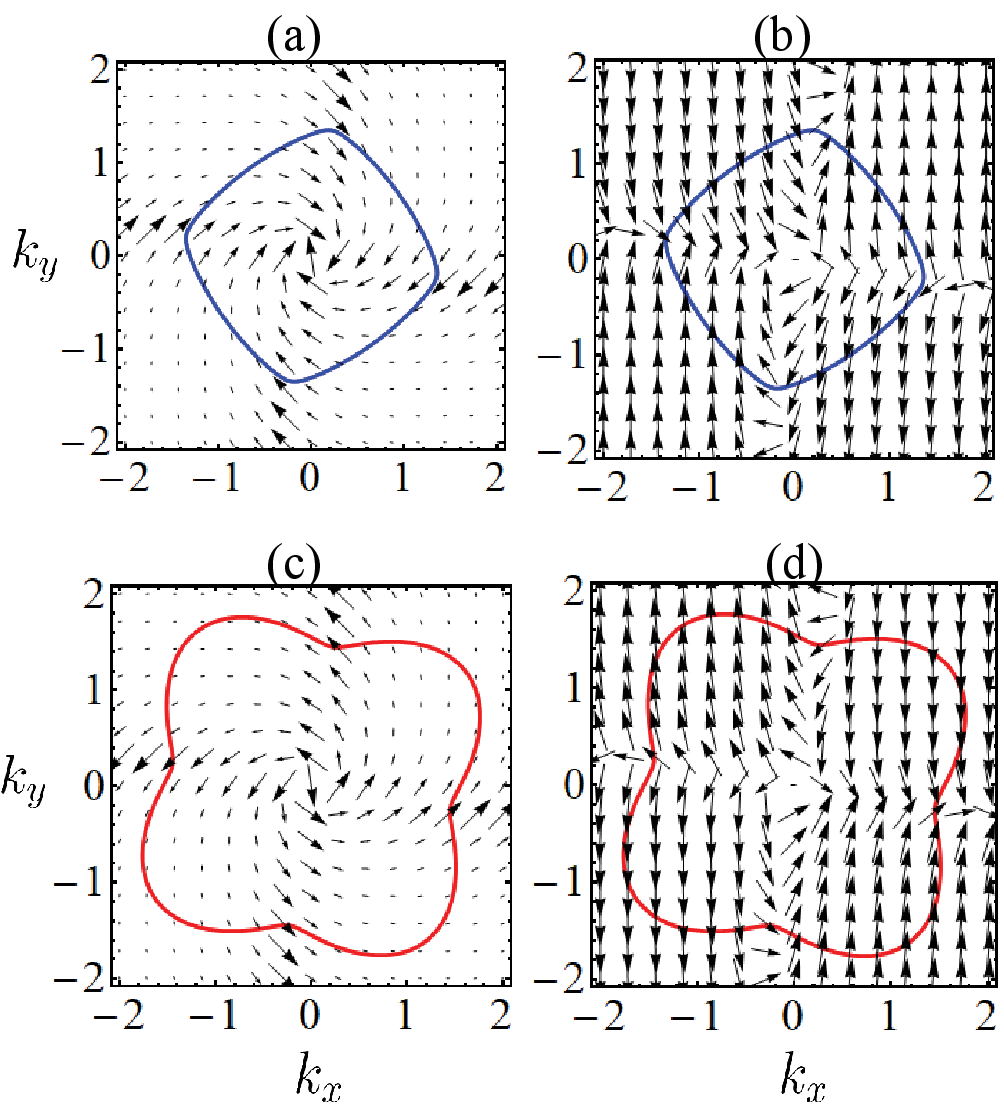}}
\caption{The Fermi surfaces and the spin structures of electrons in AM with RSOC in the momentum space. The Fermi surface denoted by the blue closed line in (a) and (b) is determined by $E_{+}=0$. The Fermi surface denoted by the red closed line in (c) and (d) is determined by $E_{-}=0$. The black arrows in (a) and (c) represent the spin vectors $(\langle s_{x+}\rangle,\langle s_{y+}\rangle)$ and $(\langle s_{x-}\rangle,\langle s_{y-}\rangle)$, respectively. The black arrows in (b) and (d) denote the spin vectors $(\langle s_{x+}\rangle,\langle s_{z+}\rangle)$ and $(\langle s_{x-}\rangle,\langle s_{z-}\rangle)$, respectively. The parameters have been taken as $t_0=1$, $t_{J}=0.5$, $\lambda=0.1$, $\alpha=0.2\pi$ and $\mu=2$.}\label{figA2}
\end{figure}

Now, we discuss the electric structure of AM with RSOC which is the normal state of AMSCs considered in our Josephson junctions.
In this situation, $t_{J}\ne 0$ and $\lambda\ne 0$. The Fermi surfaces consist of two closed curves as shown in Fig.\ref{figA2}. The blue one in Fig.\ref{figA2}(a) or (b) is determined by $E_{+}=0$ while the red one in Fig.\ref{figA2}(c) or (d) is determined by $E_{-}=0$.
The averages of the spin for particles can be calculated from Eq.(\ref{asp}).
The black arrows in Figs.\ref{figA2}(a), (b), (c), and (d)
denote the spin vectors $\langle (s_{x+}\rangle,\langle s_{y+}\rangle)$,
$\langle (s_{x+}\rangle,\langle s_{z+}\rangle)$,
$\langle (s_{x-}\rangle,\langle s_{y-}\rangle)$,
and $\langle (s_{x-}\rangle,\langle s_{z-}\rangle)$ in the whole momentum space.
From the figures, the average spin for particles on the two Fermi surfaces can be observed. Due to the coexistence of altermagnetism and RSOC,
the spin of particles will obtain both the in-plane component and the out-plane component. In other words, the three components $\langle s_{x+(-)}\rangle,\langle s_{y+(-)}\rangle$ and $\langle s_{z+(-)}\rangle$ are all non-zero.
However, the in-plane component of spin is dominant only in the zones around $\theta=\alpha+(2n+1)\pi/4$ as shown in Figs.\ref{figA2}(a) and (c) while the out-plane component is dominant in the other zones. This can also be clearly seen in Figs.\ref{figA2}(b) and (d). This is because the spin is unpolarized along the direction $\theta=\alpha+(2n+1)\pi/4$ in the AM without RSOC as discussed above and as shown in Fig.\ref{figA1}(a). In the zones around this direction, the introduction of RSOC will pin the spin in the $xy$ plane.
In the other zones, the spin is polarized along the $z$ direction in the AM before introducing RSOC. Hence, the spin will nearly keep the polarization after the introduction of a small RSOC.

\subsection{The derivation of the Green functions} \label{B}

\begin{figure}[!htb]
\centerline{\includegraphics[width=0.9\columnwidth]{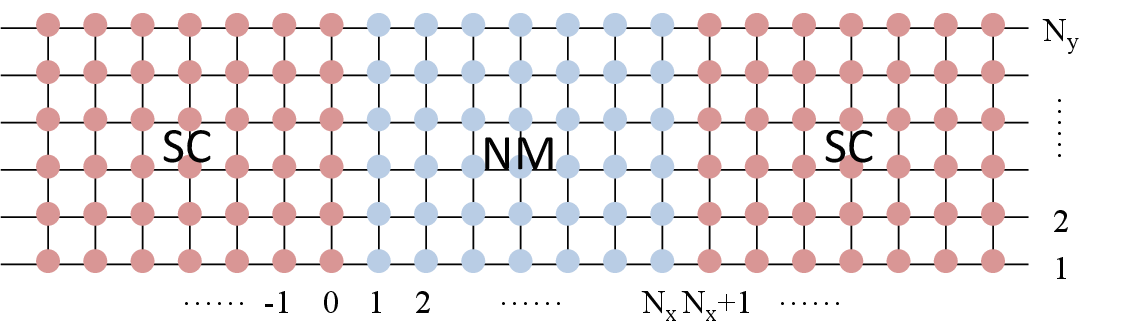}}
\caption{The two-dimensional square lattice on which the continuum Hamiltonian is discretized.}
\label{figA3}
\end{figure}

The continuum Hamiltonian in Eq.(\ref{cH}) can be discretized on the two-dimensional square lattice with the lattice constant $a$ as shown in Fig.\ref{figA3}. From the continuum Hamiltonian in Eq.(\ref{cH}), the matrices in the discrete Hamiltonian in Eq.(\ref{dH}) can be obtained. For clarity, we express the matrices $\check{H},\check{H}_{x},\check{H}_{y},\check{H}_{xy},\check{H}_{x\bar{y}}$ for the left (right) AMSC, and NM in Eq.(\ref{dH}) as $\check{H}_{L(R)},\check{H}_{L(R)x},\check{H}_{L(R)y},\check{H}_{L(R)xy},\check{H}_{L(R)x\bar{y}}$ and $\check{H}_{N},\check{H}_{Nx},\check{H}_{Ny},\check{H}_{Nxy},\check{H}_{Nx\bar{y}}$, respectively. These matrices are given by
\begin{equation}
\begin{aligned}
\check{H}_{L(R)}&=(\frac{4t_0-\mu}{a^2})\tau_z\otimes\sigma_0\\
&-\Delta_{0}[\cos{\phi_{l(r)}}\tau_y\otimes\sigma_y+\sin{\phi_{l(r)}}\tau_x\otimes\sigma_y],\label{checkH}
\end{aligned}
\end{equation}

\begin{equation}
\begin{aligned}
\check{H}_{L(R)x}&=-\frac{t_0}{a^2}\tau_{z}\otimes\sigma_0-\frac{t_J\cos{2\alpha_{l(r)}}}{a^2}\tau_z\otimes\sigma_z\\
&+\frac{i\lambda}{2a}[\cos{\alpha_{l(r)}}\tau_{z}\otimes\sigma_y+\sin{\alpha_{l(r)}\tau_0\otimes\sigma_x}],\label{Hx}
\end{aligned}
\end{equation}

\begin{equation}
\begin{aligned}
\check{H}_{L(R)y}&=-\frac{t_0}{a^2}\tau_z\otimes\sigma_0 +\frac{t_J\cos{2\alpha_{l(r)}}}{a^2}\tau_z\otimes\sigma_z\\
&-\frac{i\lambda}{2a}[\cos{\alpha_{l(r)}}\tau_0\otimes\sigma_x-\sin{\alpha_{l(r)}}\tau_z\otimes\sigma_y],\label{Hy}
\end{aligned}
\end{equation}
and
\begin{equation}
\begin{aligned}
\check{H}_{L(R)xy}=-\check{H}_{L(R)x\bar{y}}=-\frac{t_J\sin{\alpha_{l(r)}}}{2a^2}\tau_z\otimes\sigma_z.\label{Hxy}
\end{aligned}
\end{equation}
for the left (right) AMSC. For the central NM, one has
\begin{eqnarray}
\check{H}_{N}&=&(\frac{4t_0}{a^2}-\mu)\tau_{z}\otimes\sigma_0,\label{checkHn}\\
\check{H}_{Nx}&=&\check{H}_{Ny}=-\frac{t_0}{a^2}\tau_z\otimes\sigma_0.\label{Hnx}\\
\check{H}_{Nxy}&=&\check{H}_{Nx\bar{y}}=0.
\end{eqnarray}

The hopping matrices in Eq.(\ref{hH}) can be given by
\begin{eqnarray}
\check{T}_{L}=\check{T}_{R}=\text{diag}(t,t,-t,-t).\label{hH0}
\end{eqnarray}
In order to facilitate the numerical calculations, we introduce the unitary operator\cite{addr2}
\begin{eqnarray}
U=e^{\sum_{\bf{i}\sigma}[-(\frac{i\phi_{l}}{2}\psi^{+}_{L{\bf{i}}\sigma}\psi_{L{\bf{i}}\sigma}
+\frac{i\phi_{r}}{2}\psi^{+}_{R{\bf{i}}\sigma}\psi_{R{\bf{i}}\sigma})]},
\end{eqnarray}
which can lead to the transformations $U\psi_{L(R){\bf{i}}\sigma}U^{-1}=e^{i\phi_{l(r)}/2}\psi_{L(R){\bf{i}}\sigma}$ and $U\psi^{+}_{L(R){\bf{i}}\sigma}U^{-1}=e^{-i\phi_{l(r)}/2}\psi^{+}_{L(R){\bf{i}}\sigma}$. Then, the transformation on the discrete Hamiltonian $H$ in Eq.(\ref{dH}) will change $\check{H}_{L(R)}$ in Eq.(\ref{checkH}) into
\begin{eqnarray}
\check{H}_{L(R)}=(\frac{4t_0-\mu}{a^2})\tau_z\otimes\sigma_0-\Delta_{0}\tau_y\otimes\sigma_y,\label{checkHp}
\end{eqnarray}
and will change the hopping matrices in Eq.(\ref{hH0}) into
\begin{equation}
\begin{aligned}
\check{T}_{L(R)}=\text{diag}(te^{-i\phi_{l(r)}/2},te^{-i\phi_{l(r)}/2},-te^{i\phi_{l(r)}/2},-te^{i\phi_l(r)/2}),
\end{aligned}
\end{equation}
which is the one given by Eq.({\ref{hH}).
The other matrices in Eq.(\ref{dH}) are not changed. In other words, the expression of $\check{H}_{L(R)}$ in Eq.(\ref{dH}) is the one in Eq.(\ref{checkHp}) while the expressions of $\check{H}_{x}$, $\check{H}_{y}$, $\check{H}_{xy}$ and $\check{H}_{x\bar{y}}$ for AMSCs and NM in Eq.(\ref{dH}) are given by Eqs.(\ref{Hx})-(\ref{Hnx}).

In Fig.{\ref{figA3}}, the junctions have been discretized into a series of columns. Each column contains $N_{y}$ lattice points. The Hamiltonian for an isolated column in the left (right) AMSC can be given by
\begin{equation}
\begin{split}
&H_{L(R)11}=\\
&\left(\begin{array}{cccccc}
\check{H}_{L(R)}&\check{H}_{L(R)y}^{+}& & & & \\
\check{H}_{L(R)y}&\check{H}_{L(R)}&\check{H}_{L(R)y}^{+}& & &\\
 &\check{H}_{L(R)y}&\check{H}_{L(R)}&\check{H}_{L(R)y}^{+}& & \\
 & & &\ddots& &\\
 & & & & &\check{H}_{L(R)y}^{+}\\
 & & & &\check{H}_{L(R)y}&\check{H}_{L(R)}
\end{array}\right).
\end{split}
\end{equation}
The Hamiltonian for the hopping from one column to its right neighbor column is given by
\begin{equation}
\begin{split}
&H_{L(R)12}=\\
&\left(\begin{array}{cccccc}
\check{H}_{L(R)x}&\check{H}_{L(R)x\bar{y}}& & & \\
\check{H}_{L(R)xy}&\check{H}_{L(R)x}&\check{H}_{L(R)x\bar{y}}& & \\
 & \check{H}_{L(R)xy}&\check{H}_{L(R)x}&\check{H}_{L(R)x\bar{y}}& \\
  & & &\ddots& &\\
 & & & & &\check{H}_{L(R)x\bar{y}}\\
 & & & &\check{H}_{L(R)xy}&\check{H}_{L(R)x}
\end{array}\right).
\end{split}
\end{equation}
The Hamiltonian for the hopping from one column to its left neighbor column is given by $H_{L(R)21}=H_{L(R)12}^{+}$.

For NM, the corresponding matrices are given by
\begin{eqnarray}
H_{N11}=\left(\begin{array}{cccccc}
\check{H}_{N}&\check{H}_{Ny}^{+}& & & & \\
\check{H}_{Ny}&\check{H}_{N}&\check{H}_{Ny}^{+}& & &\\
 &\check{H}_{Ny}&\check{H}_{N}&\check{H}_{Ny}^{+}& & \\
 & & &\ddots& &\\
 & & & & &\check{H}_{Ny}^{+}\\
 & & & &\check{H}_{Ny}&\check{H}_{N}
\end{array}\right),
\end{eqnarray}
and
\begin{eqnarray}
H_{N12}=H_{N21}=1_{N_y\times N_y}\otimes H_{Nx}.
\end{eqnarray}

Below, we derive the surface Green functions of the left and the right AMSCs by constructing the M$\ddot{\text{o}}$bius transformation matrix\cite{Cheng2}. For the left AMSC, the matrix can be written as
\begin{eqnarray}
X_{L}=\left(\begin{array}{cc}
0&H_{L12}^{-1}\\
-H_{L12}^{+}&[(E+i\gamma)-H_{L11}]H_{L12}^{-1}
\end{array}\right)
\end{eqnarray}
with $\gamma$ being a small positive quantity. The matrix $X_{L}$ can be diagonalized by the matrix $U_{L}$, i.e., $U_{L}^{-1}X_{L}U_{L}=\text{diag}(\lambda_{L1},\lambda_{L2},\lambda_{L3}, \cdots)$. The eigenvalues satisfy the relation $\vert\lambda_{L1}\vert<\vert\lambda_{L2}\vert<\vert\lambda_{L3}\vert<\cdots$. If we express the matrix $U_{L}$ as
\begin{eqnarray}
U_{L}=\left(\begin{array}{cc}
U_{L11}&U_{L12}\\
U_{L21}&U_{L22}
\end{array}\right),
\end{eqnarray}
then the surface Green function for the left AMSC is given by $g_{L}^{r}=U_{L12}U_{L22}^{-1}$. For the right AMSC, the M$\ddot{\text{o}}$bius transformation matrix can be constructed as
\begin{eqnarray}
X_{R}=\left(\begin{array}{cc}
0&(H_{R12}^{+})^{-1}\\
-H_{R12}&[(E+i\gamma-H_{R11})](H_{R12}^{+})^{-1}
\end{array}\right).
\end{eqnarray}
Similarly, it can be diagonalized by the matrix $U_{R}$. The surface Green function for the right AMSC can be written as $g_{R}^{r}=U_{R12}U_{R22}^{-1}$ with the $12$ component $U_{R12}$ and $22$ component $U_{R22}$ of the matrix $U_{R}$.

Next, we use the recursive algorithm to calculate the retarded Green function for the leftmost column of NM. The Green function for the rightmost column of NM is given by
\begin{eqnarray}
\mathcal{G}^{r}(E,N_x)=[E-H_{N11}-\tilde{T}_{R}^{+}g_{R}^{r}(E)\tilde{T}_{R}]^{-1},
\end{eqnarray}
with $\tilde{T}_{R}=1_{N_y\times N_y}\otimes \check{T}_{R}$. The Green function for the $n$th column can be obtained by
\begin{eqnarray}
\mathcal{G}^{r}(E,n)=[E-H_{N11}-H_{N12}\mathcal{G}^{r}(E,n+1)H_{N21}]^{-1}.
\end{eqnarray}
Then, the retarded Green function for the leftmost column of NM is given by
\begin{equation}
\begin{split}
G_{N}^{r}(E)&=[E-H_{N11}-\tilde{T}_{L}^{+}g_{L}^{r}(E)\tilde{T}\\
&-H_{N12}\mathcal{G}^{r}(E,2)H_{N21}]^{-1}.
\end{split}
\end{equation} The advanced Green function for the leftmost column of NM is obtained by $G_{N}^{a}(E)=[G_{N}^{r}(E)]^{+}$.

After obtaining the surface Green functions $g_{L}^{r}(E)$ and $g_{R}^{r}(E)$, the retarded Green function $G_{N}^{r}(E)$ and the advanced Green function $G_{N}^{a}(E)$, the Green function in Eq.(\ref{JC}) can be expressed as
\begin{eqnarray}
G_{NL}^{r}(E)=G_{N}^{r}(E)\tilde{T}_{L}^{+}g_{L}^{r}(E),\\
G_{NL}^{a}(E)=G_{N}^{a}(E)\tilde{T}_{L}^{+}g_{L}^{a}(E),
\end{eqnarray}
with $g_{L}^{a}(E)=[g_{L}^{r}(E)]^{+}$. Using the fluctuation-dissipation theorem, the lesser Green function $G_{NL}^{<}(E)$ is given by
\begin{eqnarray}
G_{NL}^{<}(E)=-f(E)[G_{NL}^{r}(E)-G_{NL}^{a}(E)].
\end{eqnarray}
Substituting the expression of $G_{NL}^{<}(E)$ into Eq.(\ref{JC}), the Josephson current can be numerically calculated.

\subsection{The symmetry analysis} \label{C}

For $t_{J}=0$ and $\lambda\ne 0$, the left and the right AMSCs are time-reversal invariant due to the absence of altermagnetism. We introduce the time-reversal operation $\mathcal{T}$, which can transform the annihilation operator $c_{{\bf{k}}\sigma}$ in the following manner,
\begin{eqnarray}
\mathcal{T}c_{{\bf{k}}\sigma}\mathcal{T}^{-1}=\sigma c_{-{\bf{k}}\bar{\sigma}},
\end{eqnarray}
with $\sigma=\uparrow\downarrow$ or $\pm$. Under the time-reversal transformation, the Hamiltonian for the left (right) AMSC will become
\begin{eqnarray}
\begin{split}
\mathcal{T}H_{L(R)}(\alpha_{L(R)},\phi_{L(R)})\mathcal{T}^{-1}\\
=H_{L(R)}(\alpha_{L(R)},-\phi_{L(R)}).
\end{split}
\end{eqnarray}
Since the time-reversal operation will invert the Josephson current in the junctions, we have
\begin{eqnarray}
I(\alpha_{L},\alpha_{R},\phi)=-I(\alpha_{L},\alpha_{R},-\phi).
\end{eqnarray}
This relation means that the critical current for the positive direction and the that for the negative direction will possess the same value.
There will be no JDE for the junctions with only RSOC and without altermagnetism.

For $t_J\ne 0$ and $\lambda=0$, the time-reversal symmetry is broken due to the presence of altermagnetism. In this situation, we introduce the spin-rotation operation about the $y$ axis by a $\pi$ angle, i.e., $\mathcal{R}_y(\pi)$, which can transform the annihilation operator $c_{{\bf{k}}\sigma}$ in the following manner,
\begin{eqnarray}
\mathcal{R}_{y}(\pi)c_{{\bf{k}}\sigma}\mathcal{R}_{y}^{-1}(\pi)=\bar{\sigma}c_{{\bf{k}}\bar{\sigma}}.
\end{eqnarray}
The Hamiltonian for the left (right) AMSC is invariant under the joint operation $\mathcal{X}=\mathcal{R}_{y}(\pi)\mathcal{T}$, which leads to
\begin{eqnarray}
\begin{split}
\mathcal{X}H_{L(R)}(\alpha_{L(R)},\phi_{L(R)})\mathcal{X}^{-1}\\
=H_{L(R)}(\alpha_{L(R)},-\phi_{L(R)}).
\end{split}
\end{eqnarray}
Accordingly, the Josephson current will satisfy the following relation
\begin{eqnarray}
I(\alpha_L,\alpha_R,\phi)=-I(\alpha_L,\alpha_R,-\phi).
\end{eqnarray}
This relation implies that JDE will not exist in the junctions with only altermagnetism and without RSOC, which is irrespective of the orientation angles $\alpha_L$ and $\alpha_R$.

From the above discussions, one can find that the coexistence of RSOC and altermagnetism is the necessary condition for the JDE in our
AMSC/NM/AMSC junctions. Actually, another condition is also necessary for the JDE, which is the different orientation between the left AMSC and the right AMSC. To show this, we introduce the mirror reflection operations about $xz$ plane and the $yz$ plane, which can transform the annihilation operator $c_{{\bf{k}}\sigma}$ in the following manners,
\begin{eqnarray}
\mathcal{M}_{xz}c_{(k_x,k_y)\sigma}\mathcal{M}_{xz}^{-1}&=&\sigma c_{(k_x,-k_y)\bar{\sigma}},\\
\mathcal{M}_{yz}c_{(k_x,k_y)\sigma}\mathcal{M}_{yz}^{-1}&=&-ic_{(-k_x,k_y)\bar{\sigma}}.
\end{eqnarray}
Under these operations, the transformations of the Hamiltonian for the left (right) AMSC are given by
\begin{eqnarray}
\begin{split}
\mathcal{M}_{xz}H_{L(R)}(t_{J},\lambda,\alpha_{L(R)},\phi_{L(R)})\mathcal{M}_{xz}^{-1}\\
=H_{L(R)}(-t_{J},\lambda,-\alpha_{L(R)},\phi_{L(R)})
\end{split}
\end{eqnarray}
and
\begin{eqnarray}
\begin{split}
\mathcal{M}_{yz}H_{L(R)}(t_J,\lambda,\alpha_{L(R)},\phi_{L(R)})\mathcal{M}_{yz}^{-1}\\
=H_{R(L)}(-t_J,\lambda,-\alpha_{L(R)},\phi_{L(R)}).
\end{split}
\end{eqnarray}
As a result, the joint operation $\mathcal{Y}=\mathcal{M}_{xz}\mathcal{M}_{yz}$ will transform the Hamiltonian according to the following way,
\begin{eqnarray}
\mathcal{Y}H_{L(R)}(\alpha_{L(R)},\phi_{L(R)})\mathcal{Y}^{-1}=H_{R(L)}(\alpha_{L(R)},\phi_{L(R)}).\label{Healpha}
\end{eqnarray}
The operation $\mathcal{M}_{xz}$ will not change the Josephson current but the operation $\mathcal{M}_{yz}$ will invert the current through exchanging the superconducting phases for the left AMSC and the right AMSC. At the same time, the orientation angles in the left AMSC and the right AMSC are also exchanged. Then, we have the relation
\begin{eqnarray}
I(\alpha_L,\alpha_R,\phi)=-I(\alpha_R,\alpha_L,-\phi).\label{ealpha}
\end{eqnarray}
If one takes $\alpha_{L}=\alpha_{R}=\alpha$, the relation $I(\alpha,\phi)=-I(\alpha,-\phi)$ will hold. This means that the JDE will be absent in the AMSC/NM/AMSC junctions with the same orientation angle for the left AMSC and the right AMSC although both RSOC and altermagnetism are present. The joint operation $\mathcal{Y}$ actually represents the in-plane inversion. The relation in Eq.(\ref{Healpha}) shows that the Josephson junctions respect the in-plane inversion symmetry when AMSCs have the same orientation angle. In other words, the in-plane inversion symmetry must be broken in order to obtain the JDE in the AMSC/NM/AMSC junctions based upon AMs.

Finally, we discuss the special case of $\alpha_{L}=0$. The numerical results for this situation have been presented in the main text. We introduce the joint operation $\mathcal{Z}=\mathcal{T}\mathcal{M}_{xz}$. For the left (right) AMSC, the Hamiltonian can be transformed according to the following way
\begin{eqnarray}
\begin{split}
\mathcal{Z}H_{L(R)}(\alpha_{L(R)},\phi_{L(R)})\mathcal{Z}^{-1}\\
=H_{L(R)}(2\pi-\alpha_{L(R)},-\phi_{L(R)}).
\end{split}
\end{eqnarray}
Then, we have the relation $I(\alpha_{L},\alpha_{R},\phi)=-I(2\pi-\alpha_{L},2\pi-\alpha_{R},-\phi)$ for the Josephson current. For $\alpha_{L}=0$, the relation will degenerate into $I(\alpha_{R},\phi)=-I(2\pi-\alpha_{R},-\phi)$. This suggests that the values of the critical current for the positive direction and the negative direction will be exchanged if the orientation angle in the right AMSC is changed from $\alpha_{R}$ to $2\pi-\alpha_R$. The diode efficiency $\eta$ will change its sign correspondingly, i.e., $\eta(\alpha_{R})=-\eta(2\pi-\alpha_{R})$, which is consistent with the numerical results in Fig.\ref{fig2}(b). In addition, if $\alpha_{R}=\pi$, we can obtain $I(\pi,\phi)=-I(\pi,-\phi)$ which leads to $\eta(\pi)=0$ as shown in Fig.\ref{fig2}(b).

In the above discussions, we don't give the transformations of the Hamiltonian $H_{N}$ for NM under the time-reversal, the spin-rotation and the mirror reflection operations since $H_{N}$ will be invariant under these operations. The invariance of $H_{N}$ will not influence the above symmetry analysis about the CPRs.

\end{document}